\documentclass[pra,aps,amsfonts,amsmath,superscriptaddress,twocolumn,showpacs,floatfix]{revtex4-1}
\usepackage{graphicx} 
\usepackage{epsfig} 

\begin{document}

\title{Constraining the nuclear equation of state at subsaturation
densities}

\author{E. Khan}
\author{J. Margueron}
\affiliation{Institut de Physique Nucl\'eaire, Universit\'e Paris-Sud, IN2P3-CNRS, F-91406 Orsay Cedex, France}
\author{I. Vida\~na}
\affiliation{Centro de Física Computacional, Department of Physics, University of Coimbra, PT-3004-516 Coimbra, Portugal}

\begin{abstract}
Only one third of the nucleons in $^{208}$Pb occupy the saturation
density area. Consequently nuclear observables related to average
properties of nuclei, such as masses or radii, constrain the equation
of state (EOS) not at saturation density but rather around the
so-called crossing density, localised close to the mean value of the
density of nuclei: $\rho\simeq$0.11 fm$^{-3}$. This provides an
explanation for the empirical fact that several EOS quantities
calculated with various functionals cross at a density significantly
lower than the saturation one. The third derivative M of the energy at
the crossing density is constrained by the giant monopole resonance
(GMR) measurements in an isotopic chain rather than the
incompressibility at saturation density. The GMR measurements provide
M=1110 $\pm$ 70 MeV (6\% uncertainty), whose extrapolation gives
K$_\infty$=230 $\pm$ 40 MeV (17\% uncertainty).
\end{abstract}

\pacs{21.10.Re, 21.65.-f, 21.60.Jz}

\date{\today}

\maketitle


Constraining the nuclear equation of state (EOS) and reducing the
uncertainties on nuclear matter properties in dense stellar objects
such as neutron stars and supernovae is one of the major goals of
nuclear physics. To do so observables such as masses, radii or energy
centroid of the isoscalar Giant Monopole Resonance (GMR) are measured
in finite nuclei. Relating the EOS to such observables is usually
undertaken in several ways. The liquid drop expansion \cite{wei35} is
one of the most used methods. In its generalised version, one performs
a development of a quantity (the mass for instance) considered at
saturation density (volume term), adding several terms such as the
surface one. It should be noted that taking the volume term at
saturation density is based on the fact that the inner part of the
nuclei density is close to the saturation. Another method is the
so-called Local Density Approximation (LDA) \cite{rin80}. In this
approach the global properties in nuclei are obtained from considering
nuclei as local pieces of nuclear matter. Finally, another method is
based on the microscopic approach, relying on energy density
functionals (EDF): using various EDF's, a correlation diagram is drawn
between the predicted observable and a related property of the EOS.
The measurement of the observable allows to validate an EDF and the
corresponding property of the EOS \cite{bro00,typ01,cen09,pie11}. For
instance, the neutrons skin measurement is correlated with the slope
of the symmetry energy. It should be noted that usually
EDF are designed using masses, radii but also the nuclear
incompressibility \cite{ben03,lun03}.

Each of these methods comprise however limitations. The liquid drop
expansion is known not to be accurate enough in the case of the
incompressibility at the saturation density \cite{bla80,pea91},
although the inclusion of higher other terms is relevant \cite{pat02}.
The liquid drop expansion of the mass has been successful, providing
an accurate determination of the saturation energy. This quantity is
now used in EDF determination, but this is an exceptional case,
related to the profusion of data available on masses. In the case of
the LDA, its validity is generally questionable for finite size
systems. Finally, in the case of the microscopic approach, the EDF
employed has usually been adjusted to data on magic nuclei, whereas
most applications are aimed to be used for deformed open-shell nuclei.
However there may be a general consensus that the microscopic approach
should be used in fine, because of its accuracy and reliability.

It seems for now difficult to straightforwardly determine the nuclear
incompressibility even with the microscopic method. The earliest
microscopic analysis came to a value of K$_\infty$=210 MeV
\cite{bla80}, but with the advent of microscopic relativistic
approaches, a value of K$_\infty$=260 MeV was obtained \cite{vre03}.
It has been shown that this value is not determined accurately and
that the density dependence of the EDF as well as pairing effects (and
therefore the shell structure) have an impact on the determination of
K$_\infty$ \cite{jli08,kha09,col04}. Using K$_\infty$ in the design of
EDF may not be a sound approach since it cannot be safely determined
by the microscopic method. Therefore the method of determination of
the nuclear incompressibility has to be rethought and more generally,
it is necessary to clarify the link between nuclear matter EOS
determination and nuclear observables.

\begin{figure}[tb]
\begin{center}
\scalebox{0.35}{\includegraphics{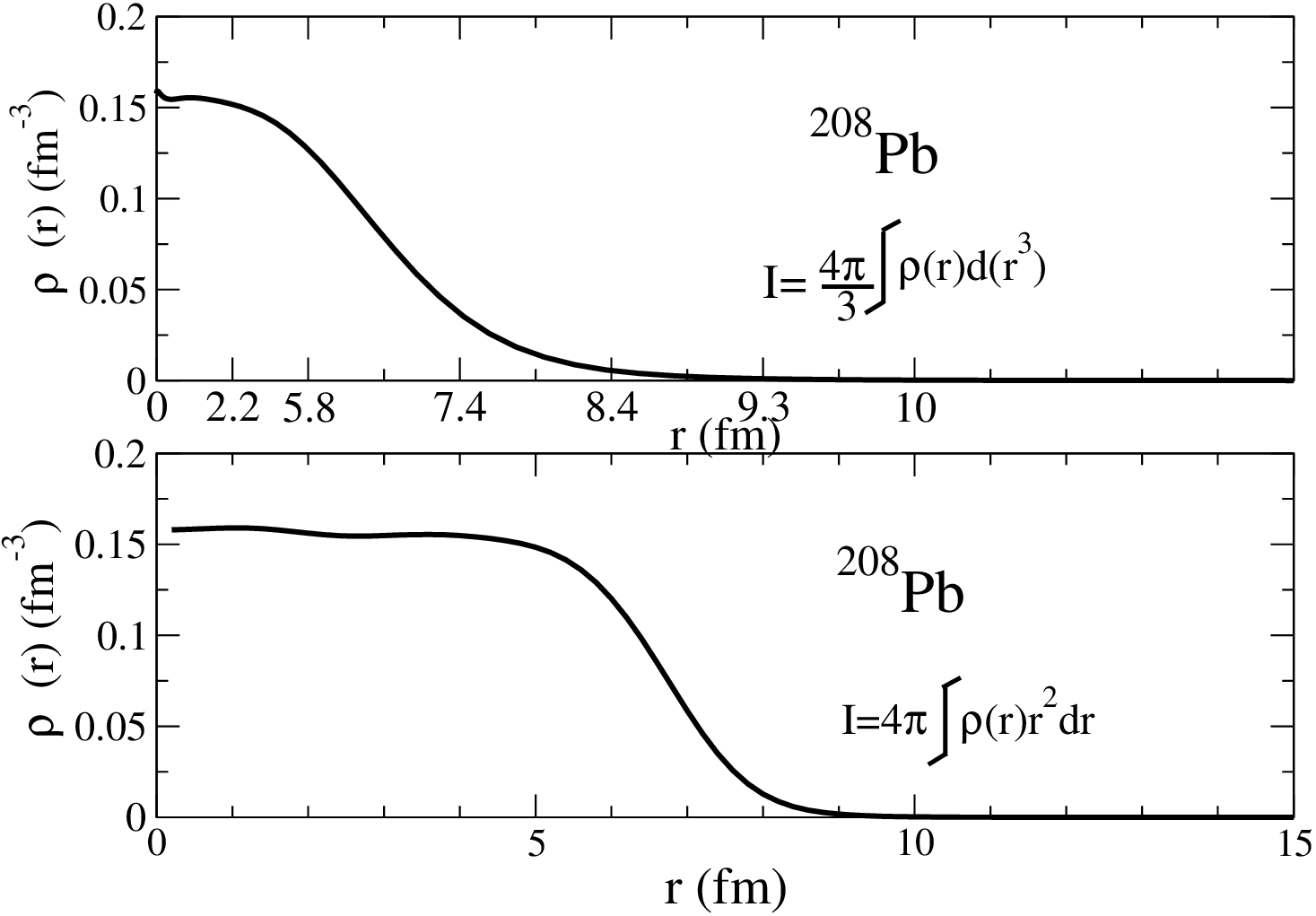}}
\caption{Matter density of $^{208}$Pb calculated with the HF approach using the SLy4
functional with different X axis scales. Top: representation taking
into account the nucleons' distribution in the nucleus. Bottom: usual representation}
\label{fig:dens}
\end{center}
\end{figure}

First it should be noted that the liquid drop expansion is not a
perturbative one since the surface properties of the nucleus are
almost as important as the bulk one. Hence it may be a misleading idea
to consider the nucleus as mainly composed of nucleons at saturation
density with a few at the surface. Let us consider the case of
$^{208}$Pb which is usually considered as a benchmark for extracting
bulk properties. Fig. \ref{fig:dens} shows the total density
calculated in the Skyrme Hartree-Fock (HF) approach. The lower part
shows the usual representation of the density whereas the upper part
displays an equivalent representation with an X axis scaled as r$^3$.
This allows to take into account the increase of nucleons per volume
unit; the total number of nucleons corresponds to the integral of
constant steps of the radial density represented on the upper part of
Fig. \ref{fig:dens}. It is now perceptible that about one third of the
nucleons of the $^{208}$Pb nucleus lie in the saturation density area,
whereas two thirds are localised in the surface at a density lower
than the saturation one. Therefore even in heavy nuclei, the
contribution of the surface is larger with respect to the volume one,
raising the question of the legitimacy of constraining EOS quantities
at saturation density by measurements of nuclear observables.

Another illustration is given by calculating the mean density of the $^{208}$Pb nucleus.
Using a Skyrme-HF calculation, one obtains
$\textless$$\rho$$\textgreater$ = 0.12 fm$^{-3}$ in $^{208}$Pb with a
variance $\sqrt{\textless \rho^2 \textgreater - \textless \rho
\textgreater^2}$=0.04 fm$^{-3}$. Therefore the density value
characterising a nucleus is not the saturation one ($\rho_0$=0.16
fm$^{-3}$) but a significantly lower one, with a range spanning from
$\rho_0$/2 to $\rho_0$. It should be noted that the dependence of the
mean density on the nucleus is rather weak: 0.11 fm$^{-3}$ for
$^{120}$Sn. In light nuclei, the mean density drops down as expected:
0.07 fm$^{-3}$ for $^{40}$Ca.

\begin{figure}[tb]
\begin{center}
\scalebox{0.35}{\includegraphics{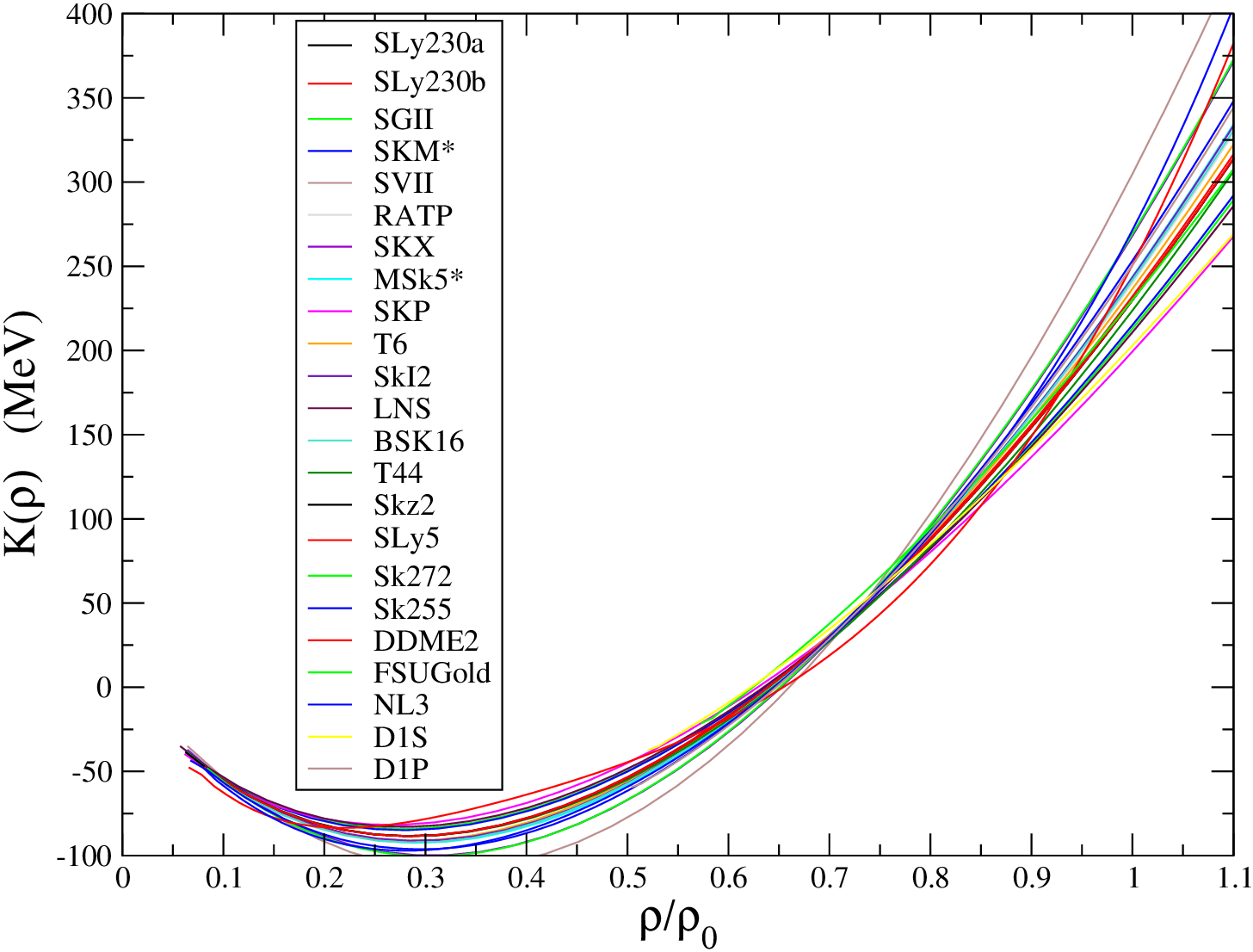}}
\caption{(Color online) EOS incompressibility calculated with
various functionals, showing the crossing density around 0.7
$\rho_0 \simeq$ 0.11 fm$^{-3}$}
\label{fig:cross}
\end{center}
\end{figure}

Consequently the measurement of an averaged observable in a nucleus 
is more properly related to a correlated EOS quantity defined around
the mean density than at the saturation density. This fact is
illustrated on Fig. \ref{fig:cross}, where the density-dependent incompressibility,
defined by \cite{fet71,kha10}

\begin{equation}
K(\rho)=\frac{9}{\rho \chi(\rho)}=9\rho^2\frac{\partial^2 E(\rho)/A}{\partial
\rho^2} + \frac{18}{\rho} P(\rho)
\label{eq:krho}
\end{equation}

and obtained from various EDF's is plotted with several Skyrme, Gogny
and relativistic interactions, all designed to describe observables in
nuclei such as masses and radii. They intersect around the crossing
density $\rho_c \simeq 0.7\rho_0\simeq$ 0.11 fm$^{-3}$. The existence
of a crossing density has been empirically noticed in previous works
on the symmetry energy (Fig. 2 of Ref. \cite{pie11}), pairing gap
(Fig. 2 of Ref. \cite{kha09b}) or the neutron EOS (Fig. 2 of Ref.
\cite{bro00} and Fig. 1 of Ref. \cite{typ01}), and we provide here an
explanation, related to the mean density: when designing EDF with
nuclear observables, the corresponding EOS is constrained not at the
saturation density but rather around the mean density (the crossing
density). In the case of the symmetry energy, the value at a density
$\rho\simeq$0.11 fm$^{-3}$ is taken to be around 25 MeV, a value close
to the symmetry energy coefficient of the liquid drop expansion
\cite{hor01,cen09,pie11}, as an empirical prescription. This last
value contains both a volume and surface terms, and thus represents
the symmetry energy extracted from nuclei observables. For the
incompressibility, the GMR is known to be related to the mean square
radius of the nucleus by the energy weighted sumrule \cite{rin80}. In
the design of EDF's, the considered constraint on nuclear radii induces a
constraint on the compression mode, likely explaining the crossing
around 0.7$\rho_0$ observed on Fig. \ref{fig:cross}. This shows the
universality of the crossing effect, arising from the constraints
encoded in the EDF from nuclei observables. Due to this crossing area,
a larger K$_\infty$=K($\rho_0$) value for a given EDF can be
compensated by lower values of K($\rho$) at sub-crossing densities, so
to predict a similar energy of the GMR in nuclei: the GMR centroid is
related to the integral of K($\rho$) over a large density range
\cite{kha10}. This allows to understand how EDF with different
K$_\infty$ can predict a similar energy of the GMR, as noted in
\cite{col04}.

Various EDF's shall exhibit various density dependencies around the
crossing point. At first order the derivative of the pairing gap (or
incompressibility or energy) at this point will differ between various
EDF's. Complementary measurements in nuclei are needed to characterise
these derivatives. For instance in Ref. \cite{bro00,typ01} the
derivative of the neutron EOS around $\rho_c\simeq$0.11 fm$^{-3}$ is
found to be constrained by neutron skin measurements in $^{208}$Pb.
The associated interpretation was that the 0.11 fm$^{-3}$density was
considered as the neutron saturation density in nuclei. We provide an
alternative and general view: 0.11 fm$^{-3}$ corresponds to the
crossing density, as seen on Fig. \ref{fig:cross}.


We apply this method to the determination of the nuclear
incompressibility. When the Giant Monopole Resonance is measured and
well reproduced by a given EDF, it shall therefore not be correlated
with the incompressibility of EOS at saturation density but rather
with its first derivative M around the crossing density. It should be
recalled that the crossing density exists because the EDF are
determined by including nuclear observables such as masses and radii in
their fit. On Fig. \ref{fig:cross} the crossing point at $\rho_c$ 
$\simeq$ 0.7$\rho_0$ yields K($\rho_c$) $\simeq$ 40 MeV for the incompressibility
(this is analogous to the fixed symmetry energy taken to be 25 MeV, as
discussed above). To be consistent with a generalised liquid drop
expansion \cite{duc11}, the derivative M of the incompressibility at
this point is defined by:

\begin{equation}
M=3\rho K'(\rho)|_{\rho=\rho_c}
\label{eq:edm}
\end{equation}

where K$\mathrm{'}$($\rho$) is the derivative of the incompressibility density
dependent term defined in Eq. (\ref{eq:krho}).

\begin{figure}[tb]
\begin{center}
\scalebox{0.35}{\includegraphics{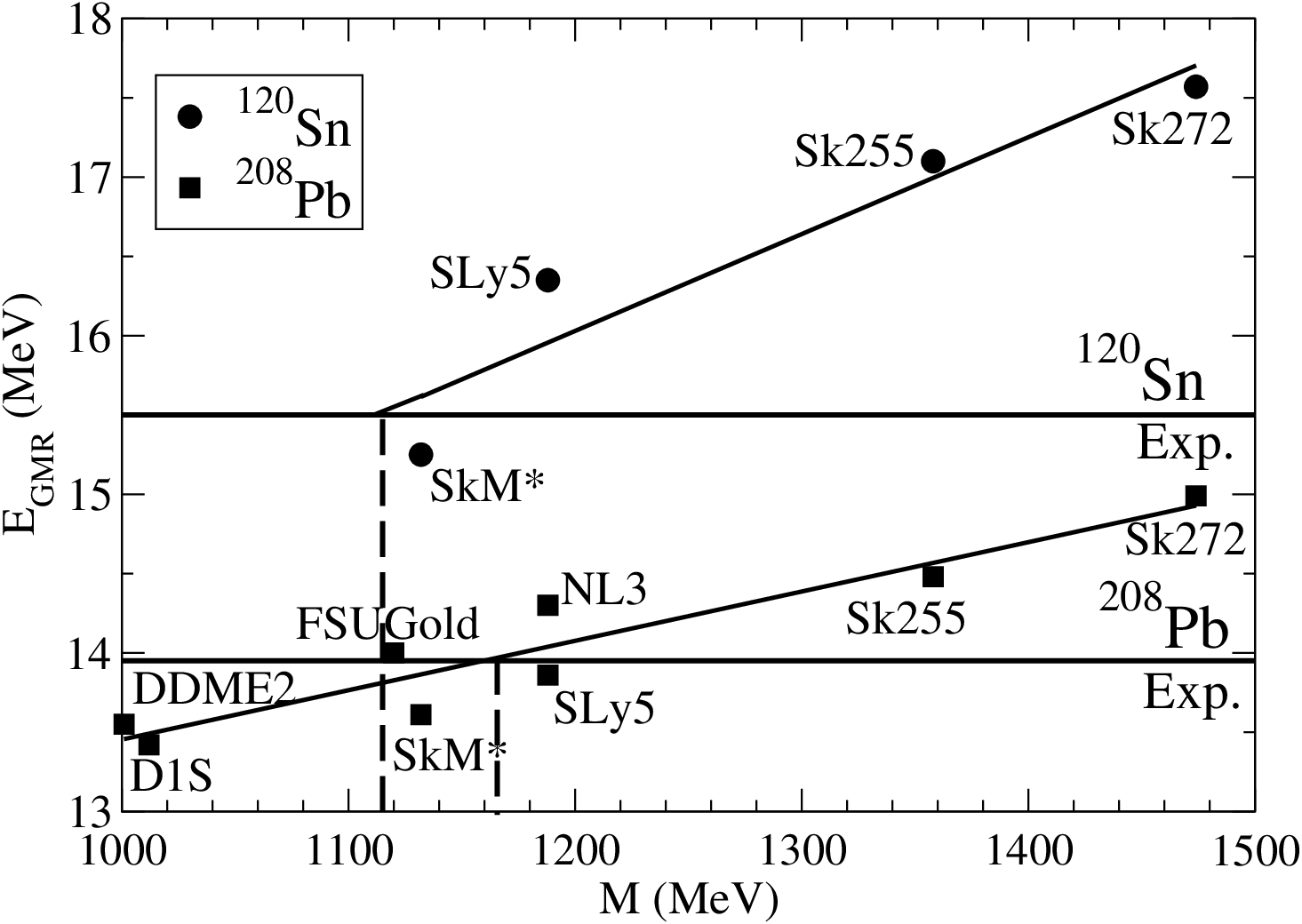}}
\caption{Centroid of the GMR in $^{208}$Pb and $^{120}$Sn calculated with the CHF method vs. 
the value of M for various functionals. The experimental values
for $^{208}$Pb and $^{120}$Sn are taken from Ref. \cite{you04} and
\cite{tli07},
respectively with respective error bars of $\pm$ 200 keV and $\pm$ 100
keV }
\label{fig:Mgmr}
\end{center}
\end{figure}

Most of the GMR measurements have been performed on $^{208}$Pb and
data on other nuclei like the Sn isotopic chain is relevant
\cite{tli07}. In Fig. \ref{fig:Mgmr} the GMR prediction using the
constrained Hartree-Fock (CHF) method as a function of M is shown for
various Skyrme EDF's in $^{208}$Pb and $^{120}$Sn. The CHF method is a
sum rule approach to calculate the centroid energy of the isoscalar
GMR:

\begin{equation}
E_{\rm ISGMR}=\sqrt{\frac{m_1}{m_{-1}}}.
\end{equation} 

The $m_1$ moment is evaluated by the double commutator using the 
Thouless theorem \cite{rin80} and for the $m_{-1}$ moment, the CHF
approach is used \cite{boh79,cap09,kha09}: the CHF Hamiltonian is
built by adding the constraint associated with the IS monopole
operator. The CHF method has the advantage to very precisely predict
the centroid of the GMR using the $m_{-1}$ sumrule. To be
comprehensive, Fig. \ref{fig:Mgmr} also displays values for several
relativistic and the Gogny D1S EDF's in the case of $^{208}$Pb.

The value of M deduced from both nuclei is compatible within few
percents. In order to derive a sound value of M, we have further
performed similar calculations in the $^{112-124}$Sn isotopes, as well
as in $^{90}$Zr and $^{144}$Sm nuclei, leading to M$\simeq$1100 MeV
$\pm$ 70 MeV. The contributions to the uncertainty shall come from the
small variance between the mean density in the respective nuclei
compared to the crossing density, the isospin dependence of the
incompressibility and the pairing effects. The linear correlation
observed on Fig. \ref{fig:Mgmr} is striking, since the interactions
employed can have a symmetry energy spanning from 30 to 37 MeV. This
shows that the present results moderately depend on the isospin
asymmetry of the system studied. The neutron vs. proton asymmetry
parameter $\delta$=(N-Z)/A also remains rather constant in the
considered nuclei for the GMR such as $^{208}$Pb and stable Tin
isotopes: $\delta \simeq 0.2$, validating the present isospin
independent approach.

It should be noted that the present method clarifies the issue of
determining different incompressibility values at saturation when
using either the Pb or the Sn data \cite{pie07,kha09}: a close M value
is found with these two data sets. However the remaining discrepancy
of the M values deduced from the Sn and the Pb measurements shows that
the proper density dependence of the EDF's has not been revealed yet.
Another striking feature is that the Gogny and relativistic EDF's are
found on the same linear correlation than the Skyrme one, showing the
universality of the E$_{GMR}$ vs. M correlation, contrary to the
E$_{GMR}$ vs. K$_\infty$ one: it is well-known that relativistic EDF's
can predict a similar E$_{GMR}$ in $^{208}$Pb but with a significantly
larger value of K$_\infty$ \cite{vre03}.

\begin{figure}[tb]
\begin{center}
\scalebox{0.35}{\includegraphics{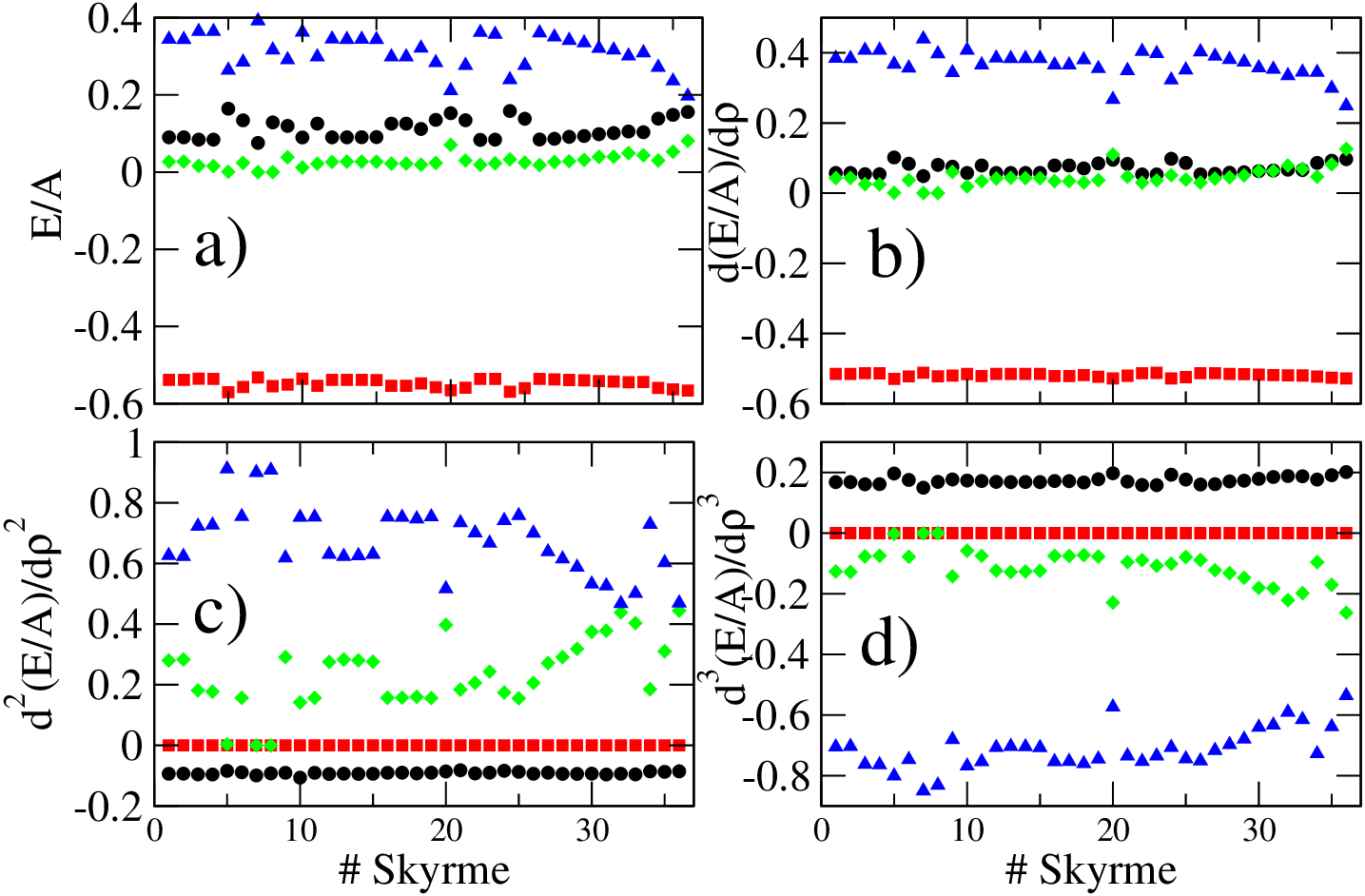}}
\caption{(Color online) Positive or negative relative contribution of the kinetic (circles), central
(squares), finite range (diamonds) and density (triangles) terms to the 
equation of state (a), its first (b), second (c) and third (d)
derivatives at the crossing density, calculated for 36 Skyrme
functionals}
\label{fig:ea}
\end{center}
\end{figure}

Let us now investigate why a clear correlation exists between the
centroid of the GMR and K$_{\infty}$ in the specific case of the
Skyrme functionals. \cite{col04}. Fig. \ref{fig:ea} displays the
relative contribution of the kinetic, central (t$_0$), finite range
(t$_1$,t$_2$) and density (t$_3$) terms of the Skyrme functionals to
the EOS and its derivative values at the crossing density. A striking
regularity is observed among the functionals used. The dominant terms
are the central and the density ones, but the central term vanishes
from the second derivative of the EOS and beyond, allowing the density
term (in $\rho^\alpha$) to dominate alone. Therefore the second
derivative terms and beyond are correlated to each other, implying
that the correlation between M and E$_{GMR}$ is propagated to the one
between K$_{\infty}$ and E$_{GMR}$: a linear correlation is preserved
between M and K$_\infty$ due to the vanishing of the central term from
the second derivative and beyond of the equation of state. In other
words the density dependence of the incompressibility, driven by these
derivatives, is correlated to M. Therefore K$_{\infty}$ is correlated
to M, and this outcome is similar for all Skyrme functionals as seen
on Fig. \ref{fig:ea}: the (E$_{GMR}$,K$_\infty$) linear correlation
may be due to an artefact from the specific density dependence of the
Skyrme EDF's.

But this linear correlation vanishes when including relativistic EDF
predictions, as mentioned above. M shall be the exclusive quantity
which can be deduced from GMR measurements, whereas the K$_\infty$
determination relies on additional density dependence assumptions.
Using the M value at the crossing point, a linear expansion provides
K$_\infty$=230 MeV. An uncertainty of $\pm$ 40 MeV can be inferred
from the spreading of K$_\infty$ values on Fig. \ref{fig:cross}
obtained with the various functionals, which is of the order of $\pm$
17\%. It is therefore argued that measurements in nuclei constrain the EOS
around the crossing density (namely the derivative of the EOS quantity
at the crossing density) and deducing values at saturation density
remains a mainly model-dependent extrapolation. In the case of
nuclear incompressibility, it is proposed to change the usual
E$\mathrm{_{GMR}}$ vs. K$_{\infty}$ correlation plot (only valid in
the specific density dependence of Skyrme EDF's) and to replace it by
a more reliable and universal E$\mathrm{_{GMR}}$ vs. M plot (Fig.
\ref{fig:Mgmr}), where M is the derivative of the incompressibility at
the crossing density. The measurement of the GMR in more neutron-rich
nuclei \cite{mon08,van10} will certainly open the possibility to study
a part of the isospin dependence of the incompressibility.


This work has been partly supported by the ANR NExEN and SN2NS
contracts, the Institut Universitaire de France, by CompStar, a
Research Networking Programme of the European Science Foundation and
by the initiative QREN financed by the UE/FEDER through the Programme
COMPETE under the projects, PTDC/FIS/113292/2009, CERN/FP/109316/2009
and CERN/FP/116366/2010. J.M. would like to thank K. Bennaceur and J.
Meyer for useful discussions during the completion of this work.

\end{document}